\documentclass[12pt,preprint]{aastex}

\newcommand\simlt{\lower.5ex\hbox{$\; \buildrel < \over \sim \;$}}
\newcommand\simgt{\lower.5ex\hbox{$\; \buildrel > \over \sim \;$}}

\begin{document}

\title{On the Origin of Rapid Flares in TeV Blazars}
\author{Amir Levinson \altaffilmark{1}}
\altaffiltext{1}{School of Physics \& Astronomy, Tel Aviv University,
Tel Aviv 69978, Israel; Levinson@wise.tau.ac.il}

\begin{abstract}
The rapid variability of the VHE emission reported for some TeV blazars implies Doppler factors
well in excess of those inferred from superluminal motions and unification schemes.
We propose that those extreme flares may result from radiative deceleration of blobs 
on scales where local dissipation occurs.  The minimum jet power estimated from the resolved synchrotron 
emission on VLBI scales appears to be consistent with this model.
It is shown that if the energy distribution of nonthermal electrons accelerated locally in the blob is reasonably flat, then
a background radiation field having a luminosity in the range 10$^{41}$-10$^{42}$ erg s$^{-1}$ can give rise to 
a substantial deceleration of the blob, but still be transparent enough to allow the TeV $\gamma$-rays thereby produced
to escape the system.
\end{abstract}


\section{Introduction}
Very high energy emission (VHE, $E\simgt100$ GeV) has been detected from over a dozen blazars \citep[for an updated
list see, e.g.,][]{wag07}, all of which are exclusively associated with the class of high peak BL Lac 
objects.  The observed bolometric luminosity during quiescent states is typically
of the order of a few times $10^{44}$ ergs s$^{-1}$, with about 10 percents emitted as VHE $\gamma$-rays. 
The luminosity in the VHE band may be larger by a factor of 10 to 100 during flaring states.
 The intrinsic spectra (corrected for absorption on the extragalactic background light) appear to be rather steep, but seem to harden  
with increasing flux.  A peak photon energy in excess of 10 TeV has been measured in the most extreme cases.  Deviations from a power law 
are clearly seen in the data in some cases \citep[e.g.,][]{alb07}. 

Rapid variability is a characteristic property of VHE blazars.  Large amplitude
variations of the VHE $\gamma$-ray flux on time scales of several hours and less have been reported for 
Mrk 421, Mrk 501 and PKS 2155-304.  This rapid variability implies  that the emission at the observed energy 
originated from a region of size (as measured in the Lab frame)
$\Delta r\simlt10^{14}\Gamma D t_{\rm var,h}/(1+z)$ cm, where $\Gamma$ and $D$ are the bulk Lorentz factor 
and the corresponding Doppler factor of the emitting matter, respectively, $t_{\rm var,h}$ is the observed variability 
time in hours, and $z$ is the redshift of the source.  If the emission originated from small jet radii, 
$r_{\rm em}\sim \Delta r$, then the requirement that the $\gamma$ rays will not be absorbed by pair production on local 
synchrotron photons implies high Doppler factors, $D\sim 30 -100$ \citep{kra02,lev06,beg07}.  
Such high values are consistent with those obtained from fits of the SED to a homogeneous SSC model\footnote{Compactness 
would constrain the location of the TeV emission region in this model}, but are in clear disagreement with the much lower 
values inferred from unification schemes \citep{Urry91,Hardcastle03} and superluminal motions on parsec scales \citep{mar99,Jorstad01,Giroletti04}.  
Various explanations, including a structure consisting of interacting spine and sheath \citep{Ghisellini05}, opening 
angle effects \citep{Gopal-Krishna04} and jet deceleration \citep{Geo03,Piner05} have been proposed in order to resolve 
this discrepancy.  A variant of the decelerating jet model is considered further below. 

It has been argued that such high values of the Doppler factor may not be required if the 
$\gamma$-ray production zone is located far from the black hole, at radii $r_{\rm em}>>\Delta r$.
In that case the compactness of the TeV emission zone may be constrained by the variability of the IR flux observed simultaneously
with the TeV flare, allowing low values of $D$ in cases where the variability time of the IR emission is much longer than
the duration of the TeV flare.  However, the fraction of jet energy that can be tapped for production of $\gamma$ rays 
in a region of size $\Delta r$ located at a radius $r_{\rm em}$ is $\eta\sim (\Delta r/\theta r_{em})^2$.  As a consequence, 
either the opening angle of the jet must be very small, $\theta \sim\Delta r/r_{\rm em}<<1$, or the jet power must be much larger than the luminosity of the TeV emission measured during the flare.  Another possibility is that the TeV emission is produced by a converging shock in a reconfinement nozzle, as proposed for the HST1 knot in M87 \citep{cheu07}.  It should be noted though that in M87 the X-ray and TeV luminosities, $L_{\rm TeV}\sim L_{\rm x}\simlt10^{41}$ erg s$^{-1}$ \citep{ah06,cheu07}, are much smaller than the TeV luminosity, $L_{\rm TeV}\sim 10^{44-45}$ erg s$^{-1}$, observed typically in the TeV blazars.  Estimates of the jet power in M87 
yield $L_j\simgt10^{44}$  erg s$^{-1}$ \citep[e.g.,][]{bick96,staw06}, implying a very small conversion fraction, $L_{\rm TeV}/L_j\simlt10^{-3}$.  Even with such a small conversion efficiency an opening angle $\theta<10^{-2}$ rad is required 
if the TeV emission were to originate from the HST1 knot, unless reconfinement can give rise to sufficient convergence of the jet at the location of HST1, as proposed by \citep{cheu07}.   We note that even modest radiative cooling of the shocked jet layer in a proton dominated jet will lead to such a convergence, at least in the non-relativistic case \citep{eich82}.  The effect of cooling on the collimation of relativistic jets needs to be explored, but we expect a similar behavior.  Stationary radio features observed on VLBI scales \citep[e.g.,][]{Jorstad01} seem to indicate that recollimation shocks may be an important dissipation channel in blazsrs, and this may apply also to other sources, e.g., GRBs \citep{brom07}.  Whether the extreme TeV flares observed in VHE blazars can be accounted for by recollimation shocks at radii $r_{\rm em}>>\Delta r$ remains to be explored.  As 
stated above, this would not resolve the 'Doppler factor crises' if the IR emission will turn out to vary on timescales comparable to
the TeV emission.

\citep[][GK03]{Geo03} proposed a scenario in which the deceleration of the fast jet base is mediated by Compton scattering of synchrotron 
photons produced further downstream, in the slow part of the flow.  The major fraction of the bulk energy is
assumed to be converted to TeV radiation within the transition layer, so there is no missing energy problem in this model.  
The main motivation in that paper was to reproduce the observed SED in a source that propagates at mildly relativistic 
speeds on VLBI scales.  Using a given bulk velocity profile for the decelerating plasma GK03 computed the spectrum emitted 
from the jet and argued that deceleration from a modest Lorentz factor ($\Gamma\sim15$) down to $\Gamma\sim$ a few in a jet observed 
at sufficiently small viewing angles ($\theta_n\simlt3^\circ$) can indeed account for the average SED observed in TeV blazers.  
The dynamics of the system has not been treated in a self-consistently manner in GK03.  In particular, it has not 
been demonstrated that (i) the backward emission from the slow jet section can provide sufficient radiative drag to 
decelerate the fast jet, and (ii) that the TeV photons can escape to infinity.
Moreover, as explained above the Doppler factors inferred from opacity constraints during the extreme flaring states observed in Mrk 412, Mrk 501 and Pks 2155-304
are considerably larger than those invoked in GK03 to explain the average broad-band spectrum of these sources.

In this paper we consider the possibility that those extreme flares are produced by radiative deceleration of fluid shells
on scales where local dissipation occurs ($r_d\sim10^2-10^3 r_g$).  The dissipation may be accomplished through formation of internal shocks in a hydrodynamic jet or dissipation of magnetic energy in a Pointing flux dominated jet.
A similar model has been proposed earlier for flares in EGRET blazars \citep{rom92,lev98}.  For the TeV blazars a background luminosity $L_s\sim 10^{41}-10^{42}$ erg s$^{-1}$ would lead to a substantial deceleration of the front 
and still be transparent enough to allow the TeV $\gamma$-rays produced by Compton scattering of the background photons
to escape the system, provided the energy distribution of radiating electrons is sufficiently flat.  The ambient radiation field is most likely associated with the nuclear continuum source.  The bulk Lorentz factor of the jet during states of low activity may be appreciably smaller than that of fronts expelled during violent ejection episodes.

\section{Minimum Jet Power: A Consistency Check}
The decelerating jet scenario implies that the major fraction of the bulk energy is radiated away in the form of VHE photons
on very small scales ($\sim 10^2-10^3 r_g$).  Consequently, the remaining jet power on VLBI scales must be much smaller than
the luminosity of the VHE $\gamma$-ray emission.  As a consistency check, we estimate the minimum power of the pc scale radio jet.  
From the resolved radio synchrotron emission emitted at a radius $r=1 r_{pc}$ pc from a
jet of opening angle $\theta=0.1\theta_{-1}$ rad
we define a fiducial equipartition magnetic field strength 
$B^\ast\sim5(T_{B9}/\theta_{-1}r_{pc})^{2/7}\nu_9^{5/7}$~mG, where             
the brightness temperature $T_B=10^9T_{B9}$~K is
evaluated at the observed frequency $\nu=\nu_9$~GHz, and the radio spectral index is $\alpha_R\sim0.5$ \citep[cf.,][]{ryb79}.  
Let the jet Lorentz factor be $\Gamma$ and let its velocity makes an angle $\theta_n$ with the line of sight. 
The jet power associated with the emitting electrons and the electromagnetic 
field is then given by                                   
\begin{equation}              
L_{j}\ge{c\over2}\left({\Gamma B^\ast\theta r\over D^{5/7}}\right)^2      
\sim10^{40}\left(\frac{\theta_{-1}r_{pc}}{D}\right)^{10/7}
\Gamma^2T_{B9}^{4/7}\nu_9^{5/7}\qquad {\rm erg\; s}^{-1}   
\label{L_jmin}                                                              
\end{equation}                        
where $D=[\Gamma(1-\beta\cos\theta_n)]^{-1}$ is the Doppler factor and rough equality occurs at equipartition.

VLBA images of Mrk 421 at 15 and 22 GHz reveal a core jet morphology \citep{mar99}, with some jet components at a linear 
distance of $r_{pc}\sim1$ from the core.  The jet components contain 
only a few percents of the total flux density (of the order of 1 Jy).  The resolution at 15 GHz is about 0.5 mas.
From the above we estimate $T_{B9}(\nu=15 {\rm GHz})\simlt1$ for the jet, and a minimum power of $L_{j,\rm min}\sim10^{41}$ erg s$^{-1}$.
For Mrk 501 we obtain similar numbers.  Thus, the minimum jet power inferred from the resolved radio emission
is well below the TeV luminosity measured in these sources, $L_{\rm TeV}\simgt10^{44}$ erg s$^{-1}$.   This 
should be contrasted with FRII sources where $T_B\simgt10^{12}$ K are measured \citep{read94}, implying a minimum
jet power close to the Eddington limit, and radio jets in microquasars, where a minimum jet power in excess of the Eddington limit
has been measured on scales of $\sim10^9 r_g$ \citep{lev96,dis02}.
 
\section{Radiative Deceleration of Relativistic Shells}
Consider a fluid shell, described by a stress-energy tensor
\begin{equation}
T^{\alpha\beta} = hnu^{\alpha}u^{\beta} - p\eta^{\alpha\beta}+\frac{1}{4\pi}
(F^{\alpha\sigma}F^{\beta}_{\sigma}+\frac{1}{4}\eta^{\alpha\beta}F^2),
\label{Tmn}
\end{equation}
where $F_{\mu\nu}$ is the electromagnetic tensor,
$u^\alpha=(\Gamma,\Gamma{\bf \beta})$ is the 4-velocity of the bulk
fluid, and $n$, $p$, and $h$, are the proper particle density,
pressure and specific enthalpy, respectively, interacting with some ambient radiation field.
Suppose that the shell has been ejected and accelerated to a bulk Lorentz factor $\Gamma_0$
at some radius $r_d\sim 10^2 r_g$, at which dissipation suddenly commences, e.g., due to collision with another 
shell or with a confining medium.  The dynamics of the front is governed by the
energy momentum equations,
\begin{equation}
\frac{\partial}{\partial x^{\alpha}}T^{\alpha\mu}=S_c^{\mu},
\label{eq: MHD}
\end{equation}
where the source term $S_c^\mu$ accounts for the radiative force acting on the front.
In terms of the distribution functions of the target photons, $f_s$, and electrons (we don't distinguish here
between electrons and positrons), $f_e$, the source term associated
with the radiative force is given, in the limit of Thomson
scattering, by \citep{Sik81,Phy82}
\begin{equation}
S_c^\mu=-c\sigma_T\int{\frac{d^3p_e}{p_e^0}\int{\frac{d^3p_s}{p_s^0}f_s
f_e p_{e\alpha}p_s^\alpha\left[p_s^\mu+\frac{(p_{e\nu}p_s^\nu)p_e^\mu}{m_e^2c^2}\right]}},
\label{Sc}
\end{equation}
with $p_e^\mu$ and $p_s^\mu$ being the 4 momenta of electrons
and soft photons, respectively, as measured in the Lab frame.
We suppose that the electron distribution function is isotropic in the fluid rest frame and can be approximated as a power
law: $dn^\prime_e/d\gamma=4\pi m_e c p^2f_e(p)\propto \gamma^{-q}$;  $\gamma_{min}<\gamma<\gamma_{max}$, where $m_ec^2 \gamma$ 
is the corresponding electron energy, as measured 
in the comoving frame.  We further suppose that the photon distribution
is isotropic in the star frame.   Under the above assumptions the zeroth component of eq. (\ref{Sc}) yields,
\begin{equation}
S^0_c=-\frac{4}{3}\Gamma^3<\gamma^2>u_s\sigma_T n^\prime_{e}.
\label{S0c}
\end{equation}
Here 
\begin{equation}
n^\prime_e=\int {f_e d^3p_e}=\int_{\gamma_{min}}^{\gamma_{max}} {\frac{dn^\prime_e}{d\gamma} d\gamma},
\end{equation}
is the proper number density of nonthermal electrons,
\begin{equation}
<\gamma^2>=\frac{1}{n_e}\int {\gamma^2 f_e d^3p_e},
\end{equation}
and $u_s(r)$ is the total energy density of the target radiation field at radius $r$.
Let $u^\prime_e=\int{\gamma m_ec^2f_e d^3p_e}$ be the proper energy density of nonthermal electrons,
$<\gamma>m_ec^2=u^\prime_e/n_e$ their average energy, and define $<\gamma^2>/<\gamma>=\chi\gamma_{\rm max}$.
In terms of these quantities the radiative force term (eq. [\ref{S0c}]) can be re-expressed as
\begin{equation}
S^0_c=-\frac{4\sigma_T}{3m_ec^2}\chi \Gamma^3\gamma_{\rm max}u_su^\prime_{e}.
\label{S0c2}
\end{equation}
For the power law energy distribution invoked above we have $\chi=(2-q)/(3-q)$ if $q<2$, $\chi=[\ln(\gamma_{max}/\gamma_{min})]^{-1}$
if $q=2$, and $\chi\simeq(\gamma_{min}/\gamma_{max})^{q-2}$ if $q>2$.  With $\gamma_{min}=m_p/m_e$ and $\gamma_{max}=10^6$
we have $\chi>0.1$ for $q\le2$.

The energy flux of the decelerating front can be obtained from eq. (\ref{Tmn}):
\begin{equation}
T^{0r}=(nh+B^{\prime 2}/4\pi)\Gamma^2\beta \equiv  u_j^\prime\Gamma^2\beta,
\label{T0r}
\end{equation}
where $n$ and $h$ are defined above and $B^\prime$ is the rest frame magnetic field.
Using eqs. (\ref{eq: MHD}), (\ref{S0c}) and (\ref{T0r}) we arrive at,
\begin{equation}
\frac{d}{dr}(u^\prime_j\Gamma^2 \beta)= -\frac{4\sigma_T}{3m_ec^2}\chi \Gamma^3\gamma_{\rm max}u_su^\prime_{e}.
\label{eq-mot}
\end{equation}
For illustration suppose that $u_s(r)\propto r^{-2}$ and that the proper density and average energy of 
the nonthermal electrons are independent of radius.  In the limit $\beta=1$ the solution to 
eq. (\ref{eq-mot}) reads
\begin{equation}
\Gamma_\infty=\Gamma_0 \frac{l}{l+r_d},
\end{equation}
where $\Gamma_0=\Gamma(r=r_d)$, and the stopping length $l$ is given by
\begin{equation}
l=\frac{3m_ec^2}{2\sigma_T\chi\xi_e\Gamma_0\gamma_{\rm max}u_s(r_d)}.
\label{stoppl}
\end{equation}
In the last equation $\xi_e=u^\prime_e/u^\prime_j<1$ denotes the fraction of total jet energy carried by the 
nonthermal electron population.

In order to calculate the pair production opacity the spectrum of the target radiation field must be 
specified.  For simplicity let us assume a power law spectrum, $I_s(\epsilon_s)\propto \epsilon_s^{-\alpha}$; $\epsilon_{s,min}<\epsilon_s<\epsilon_{s,max}$, where $I_s=hcp_s^3 f_s$ is the intensity and $\epsilon_s=p_s/m_ec$ denotes
the photon energy in $m_ec^2$ units. 
The optical depth for absorption of $\gamma$ rays of dimensionless energy $\epsilon_\gamma$  by pair creation 
is then given to a good approximation by
\begin{equation}
\tau_{\gamma\gamma}(\epsilon_\gamma)\simeq \sigma_{\gamma\gamma}r_d(u_s/m_ec^2)\epsilon_{\gamma}g(\epsilon_\gamma),
\label{taugg}
\end{equation}
with $g(\epsilon_{\gamma})=(\epsilon_\gamma\epsilon_{s,min})^{\alpha-1}$ if $\alpha>1$ and  
$g(\epsilon_{\gamma})=(\epsilon_\gamma\epsilon_{s,max})^{\alpha-1}$ if $\alpha<1$, and 
$g(\epsilon_\gamma)\le1$ in both cases.  Using eqs. (\ref{stoppl}) and (\ref{taugg}), the stopping length can 
be expressed in terms of the $\gamma$-ray optical depth as,
\begin{equation}
\frac{l}{r_d}=\frac{1}{\chi\xi_e\tau_{\gamma\gamma}}\left(\frac{\sigma_{\gamma\gamma}}{\sigma_T}\right)
\left(\frac{\epsilon_\gamma}{\Gamma_0\gamma_{\rm max}}\right)g(\epsilon_\gamma).
\label{stopp2}
\end{equation}
Note that since the observed $\gamma$-rays are produced by Compton scattering of the front electrons we must have
$\Gamma_0\gamma_{\rm max}>\epsilon_\gamma$ for any $\epsilon_\gamma$.  The highest $\gamma$-ray 
energy observed is likely to be limited by opacity.  Denoting by $\epsilon_{\gamma,\rm th}$ the photon energy at which 
the pair production opacity is unity, viz., $\tau_{\gamma\gamma}(\epsilon_{\gamma,\rm th})\simeq1$, 
and taking $\sigma_{\gamma\gamma}/\sigma_T=0.2$ we find that $l/r_d<1$ if
\begin{equation}
\left(\frac{\Gamma_0\gamma_{\rm max}}{\epsilon_{\gamma,\rm th}}\right) >\frac{g(\epsilon_{\gamma,\rm th})}{5\chi\xi_e}.
\label{condition-dec}
\end{equation}
Adopting for illustration $\chi=\xi_e=0.1$, $g(\epsilon_{\gamma,\rm th})=0.1$, we conclude that extension of the 
nonthermal electron spectrum to a maximum energy $\gamma_{\rm max}$ of the order of a few times $\epsilon_{\gamma,\rm th}/\Gamma_0$
is sufficient to cause appreciable deceleration of the front.   Gamma rays having energies above  $\epsilon_{\gamma,\rm th}$ will
be degraded to somewhat lower energies by virtue of the relatively large pair production opacity.  Thus, we naively anticipate some accumulation of flux around the maximum observed $\gamma$-ray energy, at $\sim$ a few TeV, although we stress that detailed calculations are required to determine the exact shape of the spectrum.
The corresponding luminosity of the ambient radiation can be estimated from
eq. (\ref{taugg}) to be
\begin{equation}
L_s \simeq\frac{m_ec^3r_d}{\sigma_{\gamma\gamma}\epsilon_{\gamma,\rm th} g(\epsilon_{\gamma,\rm th})}\simeq
10^{40}r_{d16}(\epsilon_{\gamma,\rm th}/10^7)g^{-1}\qquad {\rm erg\; s^{-1}}.
\label{L_s}
\end{equation}
Assuming the acceleration rate of nonthermal electrons to be on the order of their gyro-frequency, we estimate 
$\gamma_{\rm max}\simlt 10^{7.5}$ for magnetic field energy density $u^\prime_B\simeq 0.1 u^\prime_j$.  For $\Gamma_0=30$
this corresponds to $\Gamma_0\gamma_{\rm max}/\epsilon_\gamma\simeq50$ for the highest $\gamma$-ray energy observed ($\sim 10$ TeV).
If the ambient photons originate from a central continuum source (e.g., accretion disk), and about 10 percents
are intercepted by the jet, then of the order of $10^{41}-10^{42}$ erg s$^{-1}$, roughly the luminosity observed in LLAGNs, 
is required to accommodate the spectrum and variability of the TeV blazars.

We consider now the possibility that the target photons originate from the slow part of the flow (GK03).  Assuming 
for illustration that the proper energy density $u_j^\prime$ remains constant we have 
$L_{j\infty}=(\Gamma_{\infty}/\Gamma_0)^2L_j$.  The fraction of synchrotron luminosity emitted backwards (into a solid angle
$\Gamma_\infty^{-1}$ in the rest frame of the slow flow), as measured in the Lab frame, is $\sim \Gamma_{\infty}^{-4}$.  
Consequently, the luminosity of target photons is at most $\Gamma_\infty^{-4}L_{j\infty}$ 
or $L_s\simlt L_j/(\Gamma_\infty \Gamma_0)^2$.  Adopting $L_j=10^{45}$ erg s$^{-1}$, roughly the observed TeV luminosity,
and $\Gamma_\infty=4$, we deduce that $\Gamma_0$ cannot be larger than 20 or so in order that $L_s$ will satisfy
the requirements imposed in eq. (\ref{L_s}).  Note also that this photon mediated transition is unstable, so that 
some additional mechanism is required to maintain the flow at low $\Gamma_\infty$ downstream.

\section{Discussion}
The rapid variability of the VHE $\gamma$-ray emission reported for some VHE blazars implies Doppler factors $D\simgt30$, much 
larger than those inferred from superluminal motions and unification schemes.  A plausible mechanism that can generate
such flares is radiative deceleration of high Lorentz factor shells.  The minimum jet power estimated from the resolved synchrotron 
emission on VLBI scales is consistent with this model.  Deceleration to $\Gamma\sim$ a few can be accomplished if a significant fraction
of the shell's bulk energy is carried in the form of nonthermal electrons with a sufficiently flat energy distribution 
($dn_e/d\gamma\propto \gamma^{-q}$; $q\le2$) that extends up to a maximum energy at which the pair production opacity 
is of the order of a few or larger.  The spectrum emitted during the flare is expected to be hard, with $\nu F_\nu$ peaking roughly at  
an energy at which the pair production opacity is unity.  Deviation from a power at the highest energies are expected, owing to attenuation by pair production and temporal effects.  Spectral curvature is indeed observed in some cases \citep{alb07}.  Correlation between the VHE $\gamma$-ray emission and emission in other bands, particularly X-ray, is naively anticipated, although detailed calculations are required to assess exact relations (e.g., time lags and amplitude ratios).   Such correlations have been reported for Mrk 421 \citep{fos04}.  The properties of VHE flares produced
by this mechanism should differ from those predicted for the pair cascade jets in powerful blazars, like 3C279, where propagation from low-to-high $\gamma$-ray energies is expected \citep{bln95}.

I thank C. Dermer for useful comments.  This research was supported by  an ISF grant for a Israeli Center for High Energy Astrophysics.

\end{document}